\documentstyle[preprint,aps]{revtex}
\begin{document}

\title{Influence of Dislocations in Thomson's Problem}

\author{ A.\ P\'erez--Garrido$^a$, M.\ J.\ W.\ Dodgson$^{b,\, c}$ and
M.\ A.\ Moore$^b$}

\address{$^{a\,}$ Departamento de F\'{\i}sica, Universidad de Murcia,
Murcia 30.071, Spain}

\address{$^{b\,}$ Theoretical Physics Group,
    Department of Physics and Astronomy,\\
   The University of Manchester, M13 9PL, UK}

\address{$^{c\,}$ Theoretische Physik, ETH-H\"onggerberg, CH-8093
  Z\"urich, Switzerland}

\maketitle

\begin{abstract}
{
We investigate Thomson's problem of charges on a sphere
as an example of a system with
complex interactions. Assuming certain symmetries
we can work with a larger number of charges than before.
We found that, when the number
of charges is large enough, the lowest energy states are not those with
the highest symmetry. As predicted previously by Dodgson and Moore,
the complex patterns in these states involve dislocation
defects which screen the strains of the twelve disclinations required to
satisfy Euler's theorem.
}
\end{abstract}
\section{INTRODUCTION}

The properties of real systems are often determined by a large number of
interacting entities. Progress in understanding such systems relies on the
astute combination of analytical and numerical tools.
This paper presents numerical studies on Thomson's problem for numbers of
particles much larger than have been studied in the past. Thomson's problem is
to find the lowest energy of $N$ Coulomb charges distributed on the surface of
a unit sphere. Although the problem is simple to state, its solution
is non-trivial, and numerical means are required to treat the interplay of
frustration with ordering \cite{EH91,AW94,PO96,MD96}.
On a local scale, the charges are best
distributed as a triangular lattice with six neighbours to each charge. Over
the whole sphere, however, Euler's theorem dictates that there must exist
twelve disclinations---charges with only five nearest neighbours.

The problem of constructing a lattice-like structure over a spherical surface
has attracted attention in other contexts. A model with a spherical geometry
of a two-dimensional electron system in a perpendicular magnetic field
has been used to study the quantum Hall effect \cite{H83} and vortices
in a thin-film superconductor \cite{OM92,D96,DM96}.
The magnetic field is due to a Dirac
monopole at the centre of the sphere. Using numerical work on this model
for vortices,
along with the continuum elasticity theory of topological defects
on a curved surface (see Ref \cite{SN88}), it was claimed in Ref \cite{DM96}
that
including additional defects to the necessary twelve disclinations could lower
the energy of  configurations on the sphere. These defects are dislocations
(essentially bound pairs of fivefold and sevenfold disclinations) with a
particular orientation. This improved on the work of Ref \cite{D96}
where only the
twelve disclinations were included, with the result that the energy per
particle of a lattice on a sphere was greater than for a flat plane even in
the thermodynamic limit of $N\rightarrow\infty$, because of the long range
strains associated with the disclinations. The inclusion of dislocations
allows the same thermodynamic limit to be recovered, as they may screen these
long range strains.

The continuum elasticity theory allows the number of degrees of freedom  to be
reduced from the total number of particles to the number of disclination and
dislocation defects. Following Ref \cite{DM96}, if we consider the twelve
disclinations to be fixed at the corners of an icosahedron,
the total elastic energy $E_{\rm tot}$ of the system can be written as
\[
E_{\rm tot}=E_{12}+N_dE_d+\sum_{s=1}^{12}\sum_{d=1}^{N_d}E_{sd}
+\sum_{d=1}^{N_d}\sum_{d^\prime > d}^{N_d}E_{d^\prime d},
\]
where $E_{12}$ is the interaction and self energy of the
twelve disclinations, $E_d$ is
the self energy of a dislocation and the
disclination--dislocation and
dislocation--dislocation pairwise interactions are given by $E_{sd}$ and
 $E_{d^\prime d}$ respectively. These energies vary in a non--trivial
manner with the ratio $R/l_0\propto N^{1/2}$, where $R$ is the radius of the
sphere and $l_0$ is the mean lattice spacing.
Summing up the results of
\cite{DM96}, $E_{sd}$ increases more rapidly with $N$
than the self energy and the dislocation--dislocation energy.
Depending on the orientation, it may also be negative so that
dislocations
could lower the total energy of the system. The energy landscape of a single
dislocation has minima along the lines that join the disclinations.

The values of $N$ with the most stable ground states are those which are
compatible with icosahedral symmetry \cite{D96}. The structures of proteins
(capsomeres) in the shells (capsids) of spherical viruses are known to display
icosahedral symmetry \cite{MD93},
and it was in this context that these compatible
numbers were predicted. Each possible structure is defined by the translations
along the triangular lattice vectors between each disclination $(hf,kf)$,
which gives a ``magic'' number of $N=10Pf+2$ particles where $P=h^2 +k^2 +hk$
($h$ and $k$ are integers without common factors, $f$ is an integer,
and $Pf$ is the triangulation number denoted by $T$ in the literature)
\cite{CK62,C93}. The
same numbers and structures observed in nature
(in particular $N=72$ and $132$) were found in the numerical work of
Ref \cite{D96}, and are also found in Thomson's problem.

As Thomson's problem involves a simpler interaction than the model of vortices
over a sphere, we have chosen it for numerical studies of ground states on
a sphere with large $N$. Both the numerical and analytical results of
Ref \cite{DM96}
suggested that sizes with $N$ of order $10^3$ must be reached before
the dislocations will be energetically favourable. This explains
why they have not been seen in previous work on Thomson's problem,
which has concentrated
on finding the ground states for smaller system sizes to very high
accuracy.
We will not be so concerned with locating the absolute ground state, but
rather we will look for the general features of the low energy
configurations, especially the influence of dislocations on the total energy.

\section{NUMERICAL METHOD}

To treat systems of charges with $N\sim 10^3$, a different approach is
developed from earlier work on Thomson's problem. The highest number used in
this problem before was $N=200$ using a genetic algorithm \cite{MD96}.
To reduce the number of degrees of freedom, we imposed
certain symmetries on the system: fivefold rotational symmetry about
a given axis
and a twofold rotational symmetry about a perpendicular axis (these are
symmetries possessed by an icosahedron). To do this, we first keep two charges
fixed on the poles of the sphere. We then impose the symmetries by placing
clusters of ten charges, represented by $\{\theta_i,\phi_i\}$, at the polar
coordinates $(\theta_i,\phi_i+n\frac{2\pi}{5})$ and
$(\pi-\theta_i,-\phi_i+n\frac{2\pi}{5})$ for $n=0,1,2,3,4$.

The energy $E_T$
of the system is given by the relation
\[
E_T=E_p+\sum_i^{N_c} E_{pi}+\sum_i^{N_c}E_i+\sum^{N_c}_{i>j}E_{ji}.
\]
The first term is the potential energy of the two charges on the
poles (always equal to $\frac{1}{2}$ for a unit sphere;
the potential energy of two unit charges separated by a distance $d$
is given by $1/d$),
the first sum
is the potential energy of each cluster with the poles, the second
sum is the potential energy of the charges in the same cluster and
the last term is the energy of interaction of clusters with each other,
where ${N_c}=(N-2)/10$ is the number of clusters. This
reduction of the degrees of freedom by approximately a factor of ten allows
Thomson's problem to be tractable for  thousands of charges.

It has been found numerically that the
number of metastable states $M$ increases with $N$ as\cite{EH95}
\[
M \approx .382 \times \exp (0.0497N),
\]
although this number may be reduced somewhat by the restrictions of symmetry
we have imposed.
As with the number of charges we are considering, $M$ is
extremely high, the numerical optimization routine
with arbitrary initial conditions will only find
metastable states, and give little information on the properties of the ground
state.
To avoid useless searching of configuration space,
we distribute the charges in some strategic
initial configurations before allowing them to relax.

In the relaxation process
we employ an algorithm very similar to that used by Erber and Hockney
\cite{EH91} but
we do not simply move the charges in the direction of the electrostatic force.
Instead  each charge is moved to $m$ new places around the
original position which lie on  a cone of angle
 angle $\alpha$ whose axis is the direction of the electrostatic force and  
the position with lowest energy is selected as the new position if its energy 
is lower than the original energy. If all of the $m$ places have a higher 
energy than the original the charge is not moved.
We use $m=6$, having checked that the increase of this parameter does
not change the results.
When no charges can be moved, the angle $\alpha$ is decreased and the process 
is repeated.  The 
program finishes when the energy reduction
for every charge is less than a given number.
Decreasing the size  of that number increases the accuracy to which one knows 
the energy  of the final state.

\section{RESULTS}
First, we have calculated the energy of the systems with
4002 and 5882 charges (which
correspond to $T=400$ and $T=588$) with full icosahedral
symmetry, (see Figure 1 for the case of 5882 charges). We then perturbed the 
initial configuration
by displacing some charges on the line that join two disclinations. After
relaxation, we found that the new states have less energy than those with full
symmetry. We can observe the appearance of dislocations in Figure 2.
To understand what is happening, note that charges away
from dislocations are in a less strained environment than in the case
of the icosahedral configuration.

We have also calculated the energy for $N=2132$ and $5792$,
with a special initial
configuration. In this special configuration, we place the
dislocations in a symmetric fashion. We put $5p$ (p an integer)
dislocations around each disclinations on the lines that join
disclinations (i.e. we put a total of $12\times5p$ dislocations, see figure 3)
and let the system relax.
With  this initial configuration we obtain a new set of
pseudo--magic numbers. We call them pseudo--magic as they may not have ground
states of lower energy than other numbers, but they do have the possibility of
arranging the dislocations symmetrically.
In Figure 4 we show the final state obtained after relaxation.
We see that with these system sizes, the
disclination--dislocation interaction energy
is not clearly dominant and the repulsion
between dislocations is large enough to push the
dislocations away from the line that joins disclinations.
While the system was relaxing  the charge
configuration was observed,
and we could see how dislocations moved over these lines.
The patterns formed by the dislocations may be complex, but
 the results
predicted by Dodgson and Moore \cite{DM96} are seen to be emerging as
$N$ increases. With  larger systems we suspect that the
dislocation-disclination energy will dominate so that the lines of
dislocations
will become straighter.

\section{CONCLUSIONS}

In this work, we found that the most symmetric states are not those
with lowest energy in Thomson's problem for large system sizes.
This is because the introduction of additional topological
defects---dislocations---reduces the strain energy away from the
disclinations, and the interactions between these defects allows for complex
patterns. Our results should be valid for spherical systems with
different interactions between the particles.
 For example, if spherical viruses exist with much larger numbers of
protein units on the surface than has been found to date, similar patterns
with dislocations should be present or the surface would be forced to deviate
from that of a sphere.
Dislocations in carbon nanotubes also  play an important role as
they determine their radius and helicity, and  therefore control the
electrical and mechanical properties of these new molecules
\cite{I91,D94}.

\section*{ACKNOWLEDGEMENTS}
APG  would like to acknowledge a grant and financial support from the Direcci\'on
General de Investiga\-ci\'on Cien\-t\'{\i}\-fica y T\'ec\-nica, project number PB
93/1125.

\begin{figure}
\caption{5882 charges with full icosahedral symmetry. The disclinations,
represented by larger spots, reside
  at the corners of an icosahedron.  The energy is $E_T= 17049766.73$.}
\end{figure}

\begin{figure}
\caption{5882 charges with dislocations. The dislocations are
  fivefold-sevenfold pairs and the sevenfold centres are represented by the
  largest spots. The total energy is reduced to $E_T = 17049653.24$.}
\end{figure}

\begin{figure}
\caption{5792 charges, initial symmetric configuration with dislocations.
 $E_T = 16543582.87$.}
\end{figure}

\begin{figure}
\caption{5792 charges, final configuration. The more complex arrangement of
dislocations results in a lower total energy of $E_T =16530072.09$.}
\end{figure}

\end{document}